# Conventional Half-Heusler Alloys Advance State-of-the-Art Thermoelectric Properties

Mousumi Mitra[1], Allen Benton[2], Md Sabbir Akhanda[3], Jie Qi[1], Mona Zebarjadi[3,4], David J. Singh[5], S. Joseph Poon[1,4+]

[1]*Department of Physics, University of Virginia, Charlottesville, VA 22904*
[2]*Department of Physics & Astronomy, Clemson University, Clemson, SC 29631*
[3]*Department of Electrical Engineering, University of Virginia, Charlottesville, VA 22904*
[4]*Department of Material Science & Eng., University of Virginia, Charlottesville, VA 22904*
[5]*Department of Physics & Astronomy, University of Missouri, Columbia, MO 65211*

**Abstract**

Half-Heusler phases have garnered much attention as thermally stable and non-toxic thermoelectric materials for power conversion in the mid-to-high temperature domain. The most studied half-Heusler alloys to date utilize the refractory metals Hf, Zr, and Ti as principal components. These alloys can quite often achieve a moderate dimensionless figure of merit, ZT, near 1. Recent studies have advanced the thermoelectric performance of half-Heusler alloys by employing nanostructures and novel compositions to achieve larger ZT, reaching as high as 1.5. Herein, we report that traditional alloying techniques applied to the conventional HfZr-based half-Heusler alloys can also lead to exceptional ZT. Specifically, we present the well-studied p-type $Hf_{0.3}Zr_{0.7}CoSn_{0.3}Sb_{0.7}$ alloys, previously reported to have a ZT near 0.8, resonantly doped with less than 1 atomic percent of metallic Al on the Sn/Sb site, touting a remarkable ZT near 1.5 at 980 K. This is achieved through a significant increase in power factor, by ~65%, and a notable but appreciably smaller decrease in thermal conductivity, by ~13%, at high temperatures. These favorable thermoelectric properties are discussed in terms of a local anomaly in the density of states near the Fermi energy designed to enhance the Seebeck coefficient, as revealed by first-principles calculations, as well as the emergence of a highly heterogeneous grain structure that can scatter phonons across different length scales, effectively suppressing the lattice thermal conductivity. Consequently, the effective mass is significantly enhanced from ~ 7 to $10m_e$ within a single parabolic band model, consistent with the result from first-principles calculations. The discovery of high ZT in a commonly studied half-Heusler alloy obtained through a conventional and non-complex approach opens a new path for further discoveries in similar types of alloys. Furthermore, it is reasonable to believe that the present study will reinvigorate effort in the exploration of high thermoelectric performance in conventional alloy systems.

+ Correspondence: sjp9x@virginia.edu

# Introduction

Thermoelectric materials are essential for energy technology, particularly prospective applications in waste heat recovery and solar thermal systems. Their importance is that they enable scalable direct thermal to electrical energy conversion, which is crucial for many applications[1-3]. The limiting factor for applications is typically the conversion efficiency, which controls how much energy can be recovered from a given heat source. The efficiency of a thermoelectric (TE) material is characterized by the dimensionless figure of merit ZT, $ZT=S^2\sigma/(\kappa_L+\kappa_e)T$, where S is the temperature-dependent Seebeck coefficient, $\sigma$ is the electrical conductivity, and $\kappa_L$ is the lattice contribution of the thermal conductivity and $\kappa_e$ is the electronic contribution of thermal conductivity. Since the early years of thermoelectricity, researchers have found an increasing number of materials systems that demonstrate promising thermoelectric properties, with ZT~1[4].

Among thermoelectric (TE) materials, half-Heusler phases (space group $F\bar{4}3m$) have emerged in recent years as promising materials for large-scale thermoelectric power generation in view of several favorable material properties[5,6]. Half-Heusler (HH) alloys exhibit high power factor[3, 6-8], good thermal stability, and practically non-toxicity in comparison with other state-of-the-art thermoelectric materials. Furthermore, the materials can be produced in large quantities[5, 9]. The most studied HH alloys to date belong to the RNiSn and RCoSb types, where R represents refractory metals Hf, Zr, and Ti. Similar to other thermoelectric materials, there are two basic approaches to enhancing the ZT of half-Heusler alloys, namely by lowering thermal conductivity and by raising the power factor. These two TE properties are inter-related and it often poses a challenge to simultaneously improve them. Nevertheless, advances in materials synthesis and focused experimental investigation supported by sound physical insight and fundamental underpinning, have led to high ZT~1.5 in microstructure refined n-type RNiSn based alloys[10] and ZT~1.4-1.5 in novel p-type ZrCoBi[11], NbFeSb[12], and TaFeSb[13] based alloys. Such achievements would not have been possible without the advances in nanostructuring and composition exploration and optimization. These advances are built on the deeper understanding of the relevant physics that inspires such methods as band structure engineering[14, 15], hierarchical phonon scattering[16], nanostructure design[17] and nano-grain embedment[6, 18] contributing to the realization of high-ZT HH alloys.

Current high-ZT p-type half-Heusler alloys are based on novel compositions that exploit the physics of heavy hole band[12] and high band degeneracy as well as soft phonons[11, 13]. For the more conventional p-type half-Heusler alloys, some of us[18] leveraged nano phases of $ZrO_2$ throughout the grain boundaries of p-type $Hf_{0.3}Zr_{0.7}CoSn_{0.3}Sb_{0.7}$ to reduce the thermal conductivity, yielding a maximum ZT of ~0.8 at 900 K. The latter p-type alloy has had little advancement until Chen and Ren[8] designed optimal composition schemes in HfZrCoSnSb, showing that a ZT~1 can be achieved. On the other hand, the advancements mentioned have also inspired new ways of thinking about conventional half-Heusler thermoelectric materials. In their earlier work, Simonson et al.[19] reported direct evidence of enhanced near band edge density of states in lightly doped n-type half-Heusler alloys by exploiting the concept of resonant doping, giving rise to an enhanced Seebeck coefficient. The finding may provide an additional opportunity for improving the performance of p-type half-Heusler alloys. In this article, we report significant improvements in the thermoelectric properties of a previously reported p-type $Hf_{0.3}Zr_{0.7}CoSn_{0.3}Sb_{0.7}$ alloy upon doping with minute amounts of metallic Al electronically, very dissimilar to semimetallic Sb. Nonetheless, we find that Al atoms do enter the lattice on the Sb/Sn site and, crucially, this results in a remarkable increase in ZT from 0.8 to 1.5. This increase can be ascribed to the emergence of a sharp peak in the density of states of a similar p-type half-Heusler alloy ZrCo(Sn$_{0.3}$Sb$_{0.7}$)$_{1-x}$Al$_x$, yielding "resonant states" near the band edge. The present findings could pave the way for refocusing traditional doping techniques to yield state-of-the-art thermoelectric performance in common thermoelectric materials.

**Experimental Procedures**

Elements Hf, Zr, Co, Sn, Sb, and Al were weighted according to the nominal composition Hf$_{0.3}$Zr$_{0.7}$Co(Sn$_{0.3}$Sb$_{0.7}$)$_{1-x}$Al$_x$ (x=0, 0.005, 0.01, 0.015, 0.02). Following prior work[20], a small amount (~5%) of excess Sb was added to compensate for the evaporation of Sb during arc melting. Since the preferential doping site of Al in Hf$_{0.3}$Zr$_{0.7}$Co(Sn$_{0.3}$Sb$_{0.7}$) is not known a priori, additional alloys were synthesized to investigate the dopability of the (Hf,Zr) sublattice and Co sublattice. The elements were loaded into an arc furnace in preparation for melting under the Argon atmosphere. Because of the low concentration of Al as a dopant element, Al was pre-melted with small amounts of Hf and Zr to ensure homogeneous dissolution of Al in the final product. The ingot was re-melted twice to improve the overall homogeneity. Afterward, the ingots were pulverized into fine 10-30 µm size powders, followed by consolidation using Spark Plasma Sintering (Thermal Technologies® SPS 10-4) under an axial pressure of 50 MPa at 1073 K for 10 minutes and then at 1423 K for 5 minutes in vacuum. Details of the sample preparation were reported in previous publications with appropriate modifications.[21, 22]

The crystal structure was investigated via X-ray diffraction (XRD) technique using Cu-K$\alpha$ X-rays (1468.7 eV) on a PANalytical Empyrean Diffractometer at a 3°/min scan rate. Microstructures and elemental composition profiles of the samples were analyzed via scanning electron microscopy (SEM) using backscattered electron (BSE) imaging on an FEI Quanta 650 operating at an accelerating voltage of 15 keV, a spot size of 4 nm, and a working distance of approximately 10 mm. The microstructures of the half-Heusler alloys were characterized by electron backscatter diffraction (EBSD) using a Helios UC G4 Dual Beam FIB-SEM. The samples were mounted in nonconductive epoxy. Before EBSD, the surfaces of the samples were mechanically polished with SiC abrasive papers of grit sizes 600, 1200, 2500, and 4000. The surface treatment was continued by polishing with diamond polishing suspensions of 0.25 micrometer, followed by 0.05-micrometer colloidal silica suspension. The (EBSD) images were evaluated to plot the grain size distribution curves using the Crystal Imaging software attached to the instrument.

The temperature dependence of electrical resistivity and Seebeck coefficient were measured with a ZEM 3 (ULVAC Riko, Japan) electrical properties measurement system. Thermal diffusivity (D) was measured via the Laser Flash technique (LFA 457 MicroFlash System). Heat capacity, $C_p$, was measured via Differential Scanning Calorimetry (Netzsch DSC 404C), and material density ($\rho$) was measured using the Archimedes method. Thermal conductivity was calculated using the relation $\kappa = D\rho C_v$. The lattice component to the thermal conductivity was estimated by applying the Weidemann-Franz law to the electrical part of the total thermal conductivity equation, $\kappa_L = \kappa - \kappa_e$. The Wiedemann-Franz law represents a standard model of heat conduction through current carriers that gives $\kappa_e = L\sigma T$, where L is the Lorenz number, and $\sigma$ is the electrical conductivity. Room-temperature mobility (µ) and carrier concentration (n) was obtained via the Hall Effect using the electrical transport option (ETO) of Versalab. Samples used for the measurement were polished to a thickness of approximately 0.3 mm. A magnetic field (B) was applied perpendicular to the supplied electrical current (I). In this configuration, the resistance developed across the transverse leads ($R_\perp$) is measured while the magnetic field (B) was varied from 0.1 T to -0.1 T at constant I. The slope obtained from the $R_\perp$ vs. B plot was used to estimate the Hall coefficient ($R_H$) using the formula $R_H = R_\perp d/B$. Hall carrier concentration was calculated as $n = 1/eR_H$, and the Hall mobility was calculated as $\mu = R_H \sigma$.

## Results

### Materials Structure

The phase purity of the samples was confirmed through room temperature powder XRD. As can be seen in Figure 1, the peaks correspond well to the established pattern[18]. Phase analyses of the samples were implemented via a search-match technique on experimentally acquired peak files using the PDF4+ database and High-score plus software. We have added the standard Bragg peaks of the material in Figure 1. The nature of Bragg's sharp peaks indicates the material is highly crystalline. All the indexed peaks could be ascribed to the crystal structure of the half-Heusler ZrCoSb phase (JCPDS 54-0448). All the samples appear to be single-phase with the same MgAgAs-type crystallographic structure without any impurity phase within the detection limit of XRD.

Additional XRD measurements were performed on samples doped on the Hf/Zr site and the Co site via stoichiometric deficit on those sites. The consensus is that Al causes secondary phases and phase separation immediately. The impurity phases $Al_{16}Co_7$ and $Al_{1.4}Co_{0.6}$ have appeared in the samples, marked as star and triangle signs, which again confirms from the SEM image as discussed in the subsequent section of the supplementary materials. Several X-ray diffraction patterns can be seen in the Supplementary Materials (Figure S1).

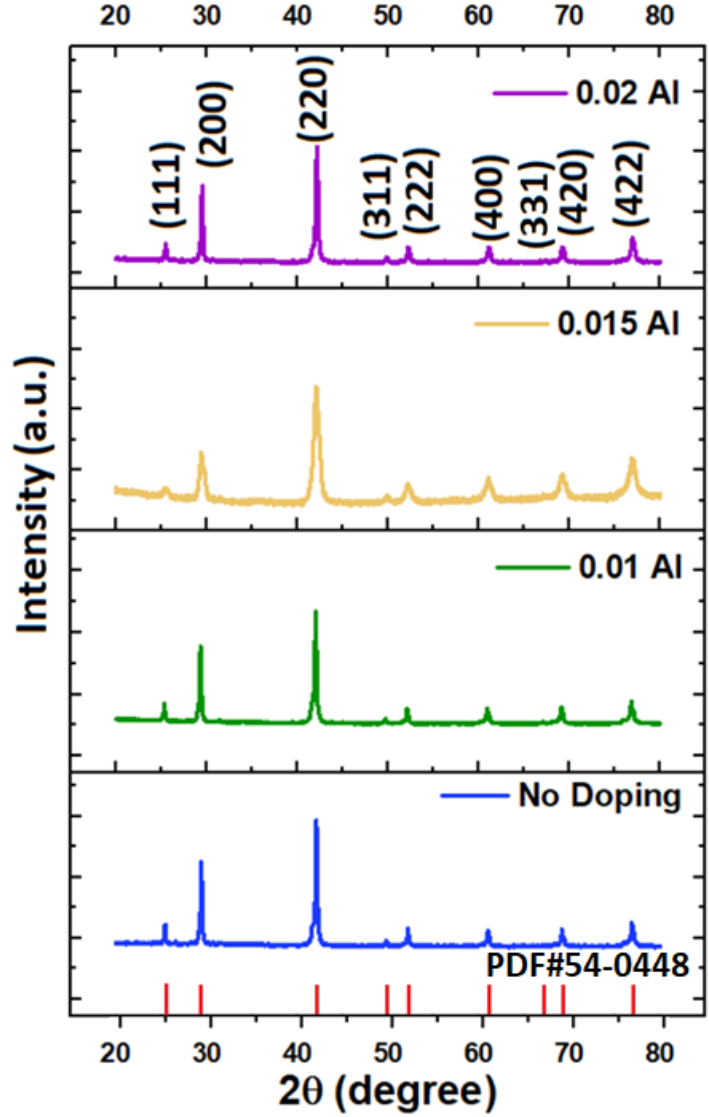

*Figure 1: Room Temperature powder X-ray diffraction patterns of $Hf_{0.3}Zr_{0.7}Co(Sn_{0.3}Sb_{0.7})_{1-x}Al_x$ where x = 0 (blue), 0.01 (green), 0.015 (yellow), 0.02 (purple).*

Since the doping ratio of Al is so tiny, any secondary phases that form would be below the resolution limit of XRD. EDS was used to obtain the compositional profile of the sample at the microscopic level. Figure 2a shows the SEM image of a specific region in the x = 0.015 sample. EDS line scan through the x = 0.015 sample shows practically no spatial variability among the elements (Figure 2b). In Figure 2c, we see general homogeneity in the elemental mapping of the x = 0.015 samples between the elements. Based on the uniform density of bright orange dots shown in the EDS mapping, Al atoms are distributed homogeneously in the sample. Results for the x = 0 (undoped), and x = 0.01 samples are shown in the supplementary materials (Figure S2). Both the undoped and x=0.01 samples also exhibit composition homogeneity. However, the x = 0.02 sample shows clear phase separation undetected by XRD. This finding indicates that no more than ~0.7 at.% Al can be doped into the Sn/Sb sublattice.

The SEM image, elemental mapping, and EDS line scan of the phase separated x = 0.02 doped on the Sn/Sb site sample are shown in the supplementary materials (Figure S3). According to EDS curve (S3b), different phases were identified and the relative amounts of all phases, with their corresponding compositions, are tabulated in supplementary Table S1.

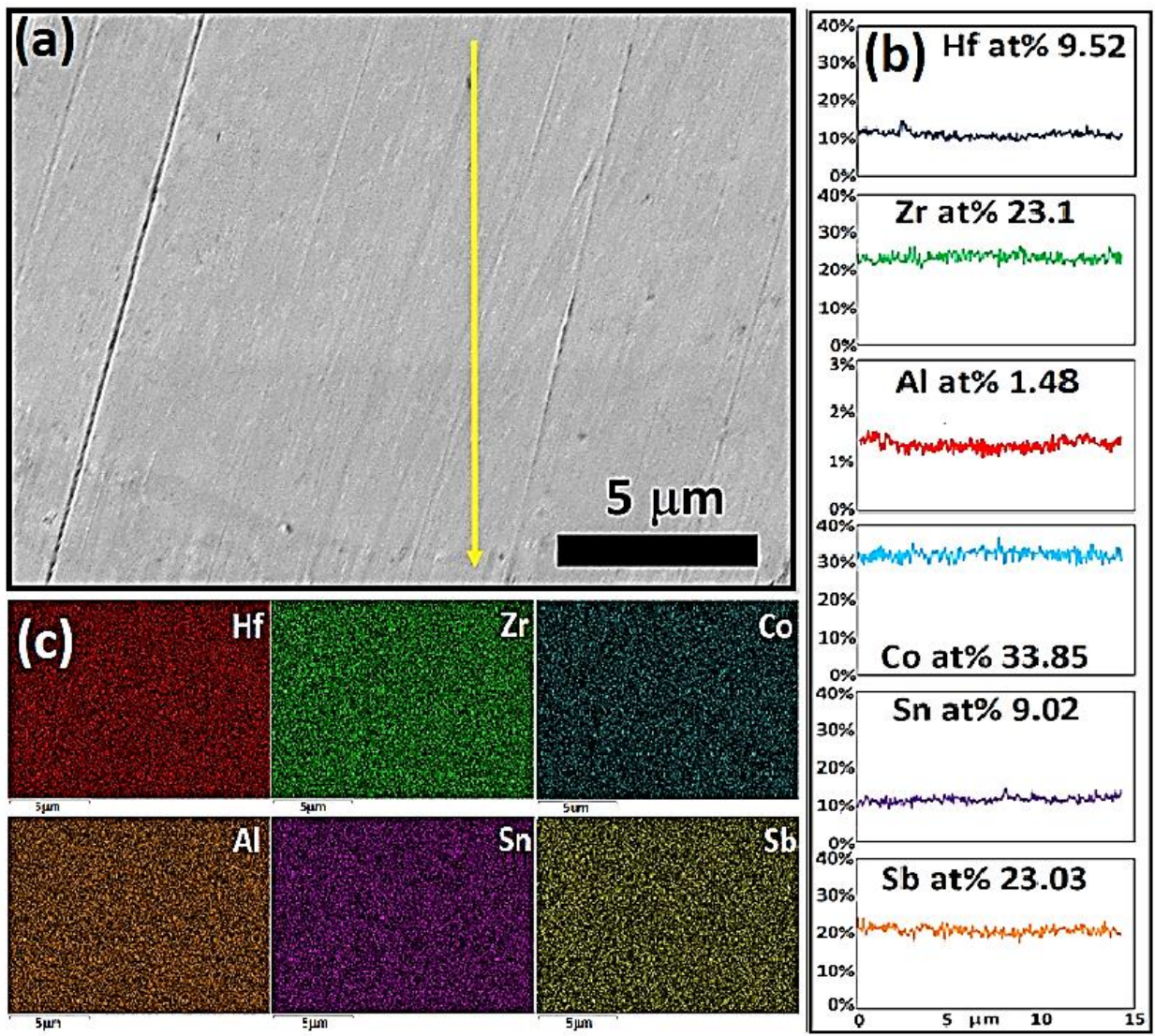

*Figure 2: (a) SEM image, (b) EDS line scan and (c) EDS elemental mapping of $Hf_{0.3}Zr_{0.7}Co(Sn_{0.3}Sb_{0.7})_{0.985}Al_{0.015}$ sample. Yellow arrow indicates the location and direction of the abscissa in the corresponding EDS curves (b).*

The Supplementary Materials (Figure S4) provide a more detailed look at these EDS scans. This figure shows $(Hf_{0.3}Zr_{0.7})_{1-x}Al_xCoSn_{0.3}Sb_{0.7}$ and $Hf_{0.3}Zr_{0.7}Co_{1-x}Al_xSn_{0.3}Sb_{0.7}$, nominally doping Al on the Hf/Zr site and the Co site. It was found that both sites are highly resistant to Al doping, and secondary phases are immediately formed. The predominant secondary phases formed are $Al_{16}Co_7$, $Al_{1.4}Co_{0.6}$, $Al_{13}Co_4$, and AlCo[23].

Figure 3 depicts the electron backscatter diffraction (EBSD) images of $Hf_{0.3}Zr_{0.7}Co(Sn_{0.3}Sb_{0.7})_{1-x}Al_x$ samples, and clear grain boundaries can be seen in these images. Starting from the undoped alloy, the grain structures of the alloys become increasingly refined as x increases. Specifically, many more micron-sized grains are formed. Surprisingly, this significant grain structure refinement can occur even at such a low level of Al doping (~0.17, 0.33, and 0.5 at. % of the alloy's total elemental content). Overall, the grain structure is observed to be highly heterogeneous. Understanding the nucleation and growth of this heterogeneous microstructure requires detailed thermodynamical and computational studies, which fall beyond the scope of the present work. The weighted-area average grain sizes for the four alloys are obtained and presented in the form of grain size histograms in Figure 4. Starting from an average grain size of ~ 3.1 μm for the host alloy, the average grain size is refined to ~ 1.6 μm for the x = 0.02 alloy.

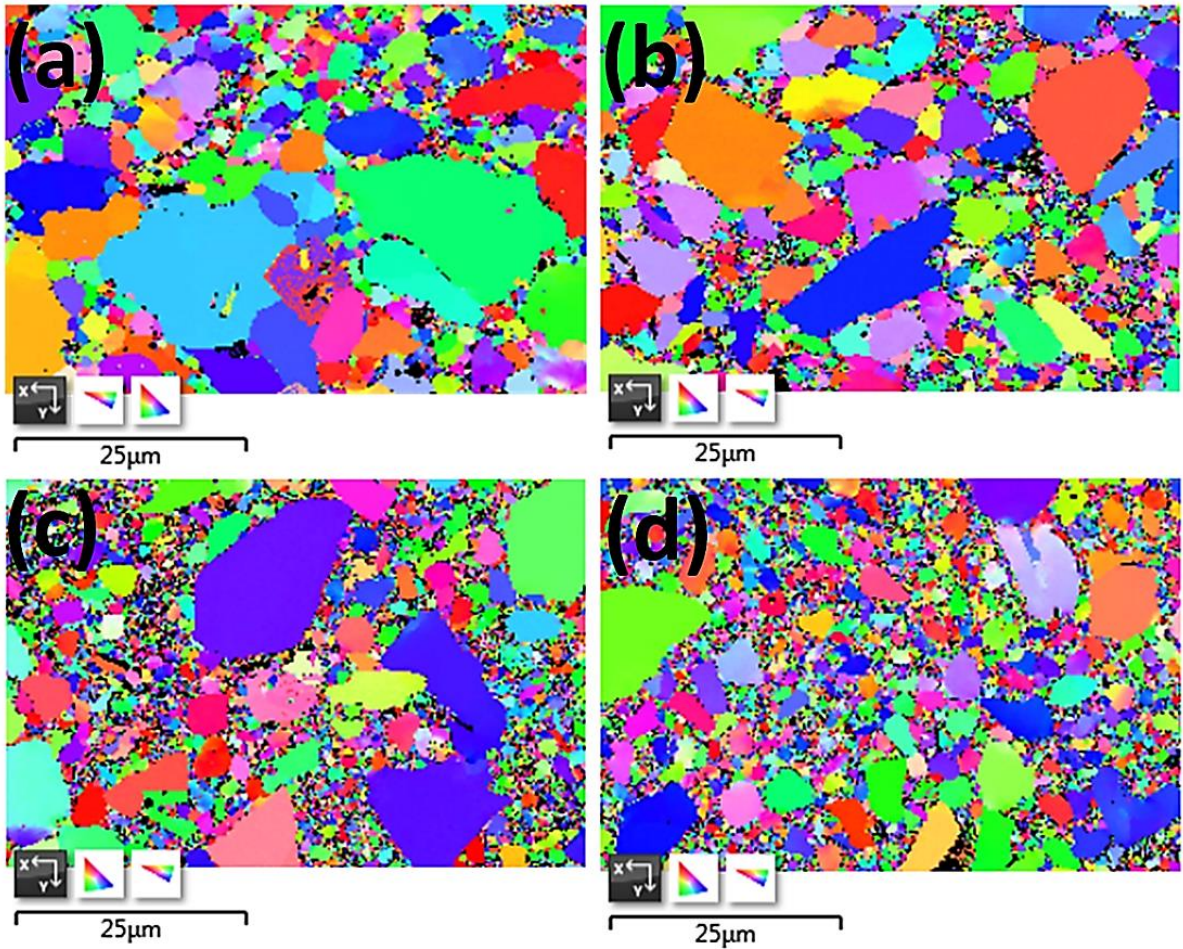

*Figure 3: Electron backscatter diffraction (EBSD) images of $Hf_{0.3}Zr_{0.7}Co(Sn_{0.3}Sb_{0.7})_{1-x}Al_x$ sample (a) x= 0, (b) x= 0.005, (c) x=0.01 and (d) x=0.015. Fine grains of micron size can be seen.*

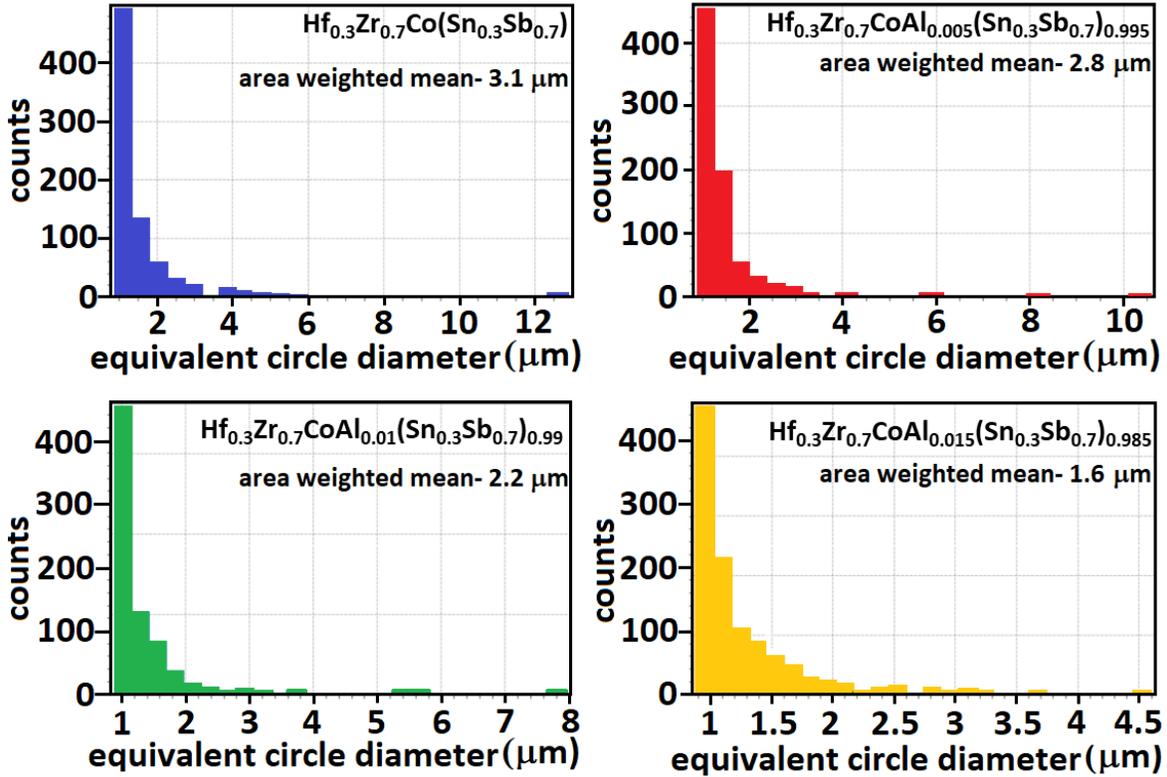

*Figure 4: Grain-size distribution histograms for $Hf_{0.3}Zr_{0.7}Co(Sn_{0.3}Sb_{0.7})_{1-x}Al_x$ (a) x=0, (b) x=0.005, (c) x=0.01 and (d) x=0.015.*

**Electrical resistivity and thermopower characterization**

The resistivity and Seebeck coefficient measured between 300 and 900K are presented in Figures 5A and 5B. There is an apparent increase in electrical resistivity as the doping ratio for Al increases until the x = 0.015 sample, which then begins to decrease as the secondary phases become more prevalent in the composition. There is a clear trend in resistivity; however, the overall variation is insignificant. The decrease in resistivity in the x = 0.02 sample indicates that the secondary phases are likely metallic.

As shown in Figure 5B, a marked increase in the Seebeck coefficient spans the entire measured temperature range for all doped samples and is trending flat in the higher temperature regions. The Seebeck coefficient systematically increases with increasing Al doping, but the change between the x = 0.01 and 0.015 samples is noticeably smaller, while the x = 0.02 sample begins to decrease slowly after 700 K.

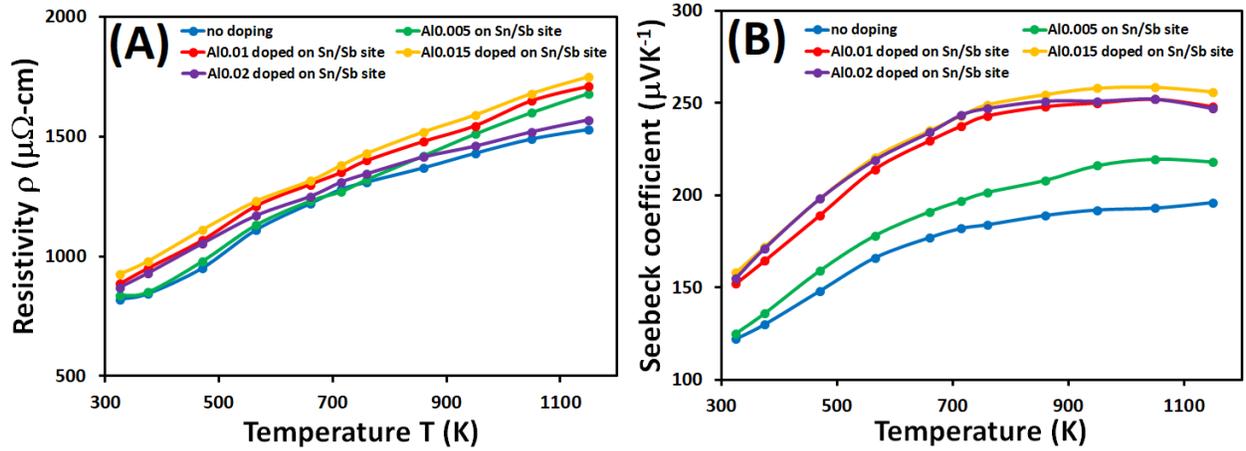

*Figure 5: Electrical resistivity of $Hf_{0.3}Zr_{0.7}Co\,(Sn_{0.3}Sb_{0.7})_{1-x}Al_x$  A) x=0, 0.005, 0.01, 0.015, and 0.02. Temperature dependent Seebeck coefficient is shown for $Hf_{0.3}Zr_{0.7}Co\,(Sn_{0.3}Sb_{0.7})_{1-x}Al_x$ (B) x=0, 0.005, 0.01, 0.015, and 0.02. (Measurements were repeated on different samples to ensure reproducibility of results).*

The combined effects on thermopower and electrical resistivity result in a considerably enhanced power factor $S^2\sigma$ compared to the undoped sample. Figure 6 highlights this enhancement of nearly 70% between the undoped and x = 0.015 samples. In comparison, the thermoelectric properties of $Hf_{0.3}Zr_{0.7}CoSn_{0.3}Sb_{0.7}$ were improved by the addition of $ZrO_2$, which formed nanoparticles in the grain boundaries, resulting in an increased power factor from 2 mW/(mK$^2$) to 2.5 mW/(mK$^2$) at 900 K[18]. At 750 K, we see a remarkable increase from 2 mW/(mK$^2$) to 4 mW/(mK$^2$), as well as a significant increase between 1 and 2 mW/(mK$^2$) can be seen along the entire measured range. Figure 6 also shows the slight increase in power factor in the x = 0.02 sample as the secondary phases become prevalent, decreasing the electrical resistivity. The electrical resistivity, Seebeck coefficient, and power factor values of the Al-doped Hf/Zr and Co site samples are presented in the supplementary materials (Figure S5).

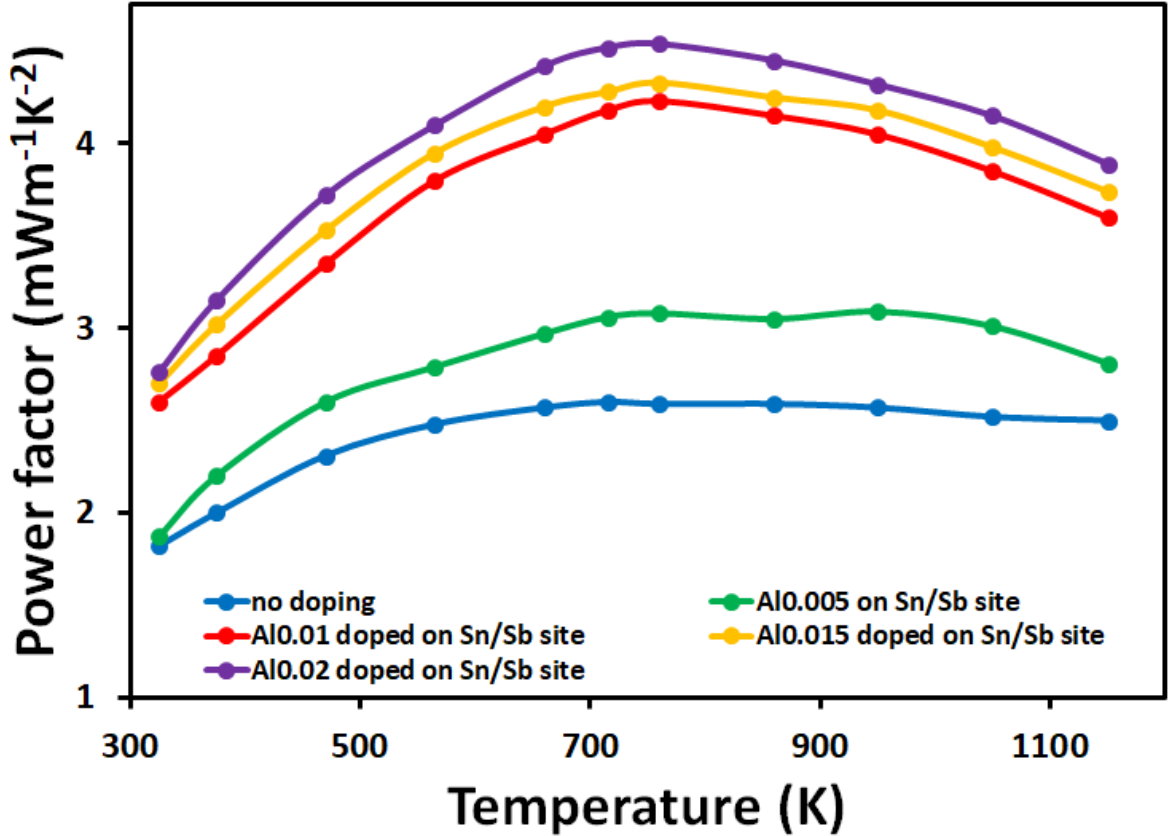

*Figure 6: Power factor for $Hf_{0.3}Zr_{0.7}Co(Sn_{0.3}Sb_{0.7})_{1-x}Al_x$ x=0, 0.005, 0.01, 0.015, and 0.02 samples.*

**Thermal characterization and Figure of merit**

The thermal conductivity from 300-1000K is shown in Figure 7. The undoped $Hf_{0.3}Zr_{0.7}CoSn_{0.3}Sb_{0.7}$ shows comparable total thermal conductivity to similar samples previously measured[18]. The doped samples show lower thermal conductivity than the undoped samples across the measured temperature range. At lower temperatures, the doped samples' thermal conductivity is notably suppressed, but the difference wanes towards higher temperatures. The lattice contribution to thermal conductivity will be discussed below.

Figure 8 shows the temperature-dependent ZT from 300 to 1000K. The undoped $Hf_{0.3}Zr_{0.7}CoSn_{0.3}Sb_{0.7}$ shows comparable ZT ~ 0.7 at 1000 K to previously reported ZT~0.6 at 1100 K[18]. A higher ZT for $Zr_{0.5}Hf_{0.5}CoSb_{0.8}Sn_{0.2}$ of 0.93 at 1123 K was also reported[24]. However, a remarkably high $ZT_{max}$ ~ 1.5 is found at 980 K for the present samples with doping levels x=0.01 and 0.015, or 0.33 and 0.5 at. % Al dopant. The plot indicates that ZT will continue to increase at higher temperatures beyond our measurement capability. At higher doping levels, ZT ceases to increase and diminishes, likely due to phase separation that attenuates the beneficial doping effect. ZT increases monotonically over the measured temperature range and shows no peak indicating higher ZT could be attained at higher temperatures. As usual, reports of thermoelectric ZT must also include uncertainty of ~ ±10%[21, 22].

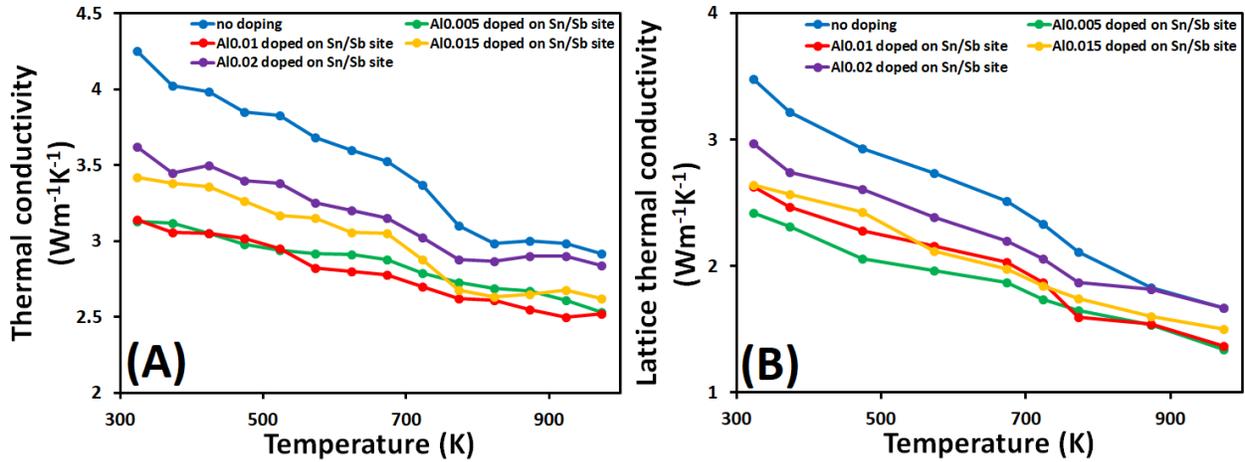

*Figure 7: Total thermal conductivity for $Hf_{0.3}Zr_{0.7}Co\ (Sn_{0.3}Sb_{0.7})_{1-x}Al_x$ (A) x=0, 0.005, 0.01, 0.015 and, 0.02. The lattice contribution to the thermal conductivity for $Hf_{0.3}Zr_{0.7}Co\ (Sn_{0.3}Sb_{0.7})_{1-x}Al_x$ (B) x=0, 0.005, 0.01, 0.015 and 0.02. (Measurements were repeated on different samples to ensure reproducibility of results).*

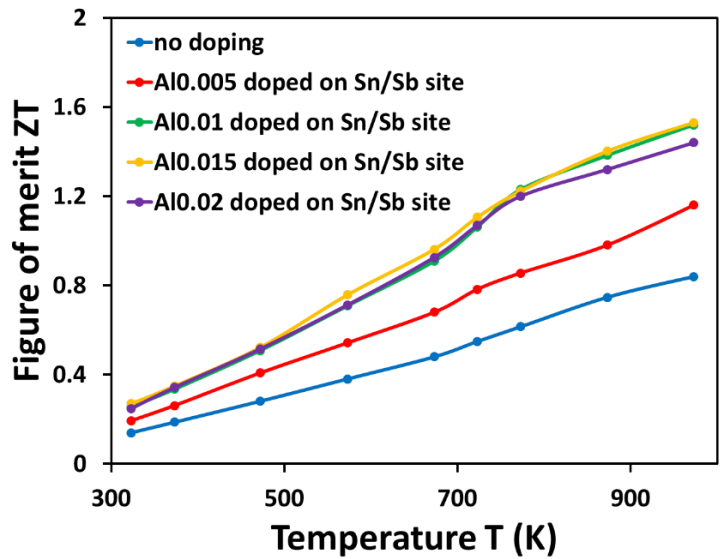

*Figure 8: Dimensionless figure of merit measured from 300-1000K is shown for $Hf_{0.3}Zr_{0.7}Co\ (Sn_{0.3}Sb_{0.7})_{1-x}Al_x$ for x= 0, 0.005, 0.01, 0.015, and 0.02.*

## First Principles Calculations

We used density functional calculations of the electronic structure to examine the effect of Al substitution on the Sn/Sb site. These calculations were done using the general potential linearized augmented plane wave method implemented in the WIEN2k code. We used the generalized gradient approximation of Perdew, Burke, and Ernzerhof (PBE GGA) and included spin-orbit for the electronic structures. We did calculations for the end-point compounds and supercells with Al replacing Sb or Sn. The lattice parameters were held fixed at the values for the corresponding pure compounds. All-atom positions within the cells were fully relaxed by total energy minimization with no imposed symmetry. Given the complex compositions of the present alloys, the calculations for the essential electronic structure were performed for the similar class of alloys ZrCoSb and ZrNiSb with Al leading to compositions $Zr_{16}Ni_{16}Sb_{15}Al$, $Zr_{32}Ni_{32}Sn_{31}Al$, $Zr_{16}Co_{16}Sb_{15}Al$, and $Zr_{32}Co_{32}Sb_{31}Al$. That is, the cells have 1/16 and 1/32 of the Sb/Sn substituted by Al, somewhat higher than for the studied alloys (≤1/50 of the Sb/Sn). For both cell sizes, we find a substantial enhancement of the density of states at the valence band edge due to hybridized Al states, and in addition, we find that Al is a p-type dopant provided one hole per Al for ZrNiSn and two holes per Al for ZrCoSb. The density of states and Al contributions are shown in Figure 9.

## Discussion

To examine the electronic and lattice contributions to thermal conductivity, we obtained the Lorenz number using the equation proposed by Kim et al.[25]. It is important to note the values of the Lorenz number fall well below the degenerate limit, as shown in Table 1. From the Wiedemann-Franz law, we calculated the electronic component of thermal conductivity ($\kappa_e$) to obtain the lattice component of thermal conductivity ($\kappa_L$), which is shown in Figure 7B. The suppression of $\kappa_L$ due to doping is essentially transferred to the suppression of total thermal conductivity. With x= 0.005 – 0.015 in the Sb/Sn sublattice, or 0.17 – 0.33 at. % Al in the crystal structure, the $\kappa_L$ suppression is unlikely due to mass fluctuation or strain effects. Rather, the suppression due to low levels of Al doping can be attributed to enhanced grain boundary scattering due to grain structure refinement that deserves future investigation.

The effects of grain boundary scattering on electrical and thermal transport can now be examined by plotting the room temperature values of $\kappa_L$ and $\sigma$ as a function of average grain size in Figure 10. Compared with the undoped sample, the x=0.005 and 0.01 samples show $\kappa_L$ decreasing by 33% and 24%, respectively, while the corresponding $\sigma$ decreases only by 2% and 7%, respectively. Since these two samples have attained the full range of power factor enhancement and thermal conductivity reduction, this suggests that grain structure has little influence on the electrical transport properties but more significant effect on the thermal conductivity, which can be reasonably understood. This can be easily understood using conventional analysis. The carrier mean-free path $\ell \approx (2m^*E_F)^{1/2}(\mu/e)$, approximated by $(2\pi^2 k_B^2 T m^*/3eS)^{1/2}(\mu/e)$, is estimated to be $\approx 1.3$ nm using the room-temperature values of S/T and parameters including effective mass m* and carrier mobility $\mu$ for the x=0.01 sample reported below. The shortness of the carrier mean-free path compared with the average grain size indicates minimal influence from grain boundary scattering. On the other hand, first-principles molecular dynamics (MD) calculation of the cumulative thermal conductivity profile in a similar half-Heusler alloy revealed that the phonons with a vast range (10 nm-10 μm) of mean-free paths have noticeable contributions to thermal conductivity[26]. Thermal transport becomes even more intriguing in the highly heterogeneous microstructure in the present alloys. The physics of phonon transport cannot be accounted for by a simple mean-field model. Heterogeneous microstructure can efficiently suppress thermal transport as it scatters phonons across different length scales in real space and wavelengths in reciprocal space[26]. Future investigations using ab initio molecular dynamics (MD) will be needed to uncover the mechanism of phonon transport in these alloys. Below, we will focus on the electronic structure origin of thermopower enhancement.

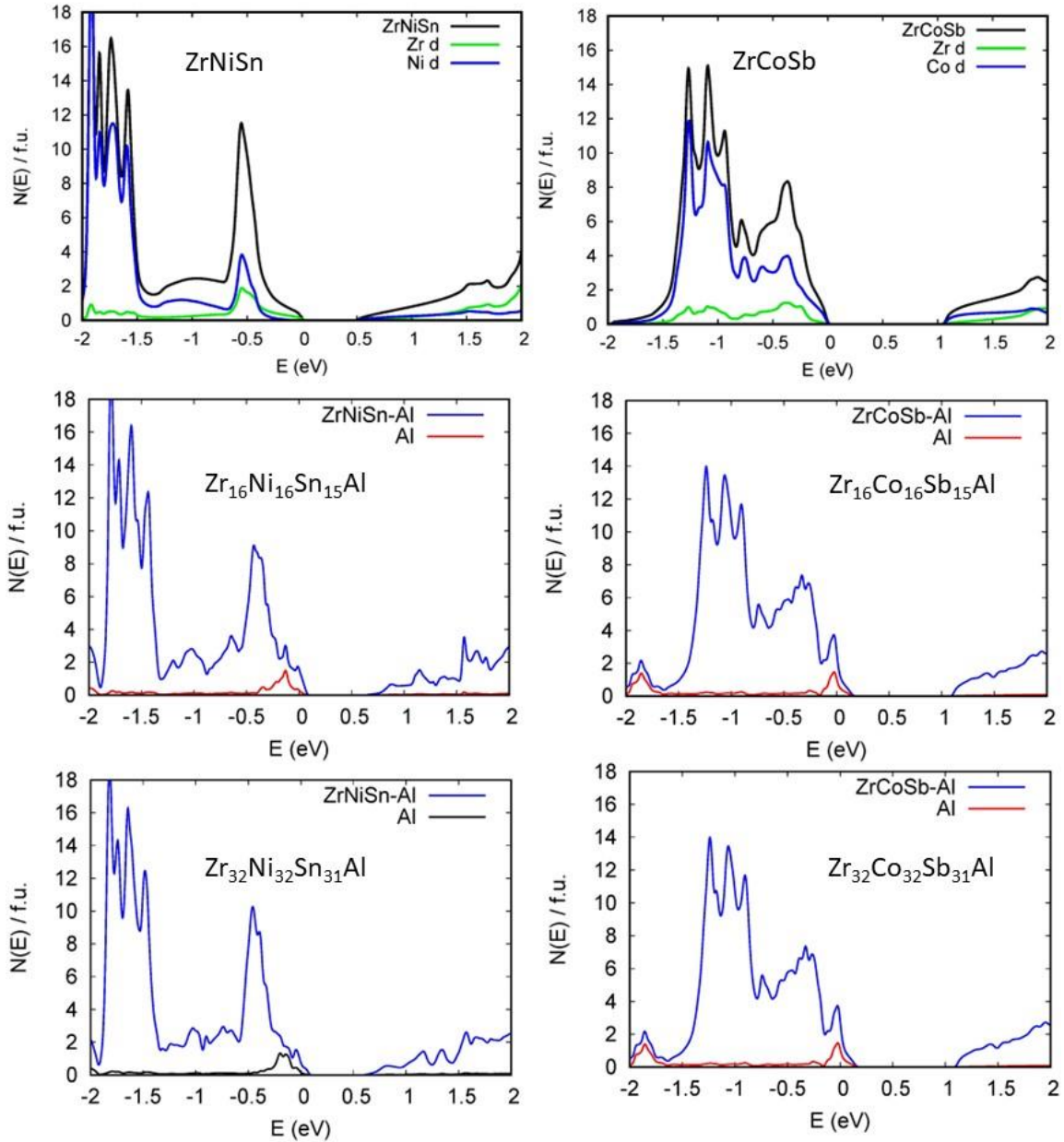

*Figure 9. Calculated electronic density of states per half-Heusler primitive cell for ZrNiSb (top left) and ZrCoSb (top right) showing projections of metal d character and supercells with 1/16 and 1/32 substitution of Al for Sn or Sb. Also shown is the projection onto the Al site in these supercells. The energy zero is at the valence band edge for the pure compound and the Fermi level position for the supercells. Note the enhancement of the DOS at the valence band edge due to hybridized Al states and the p-type doping.*

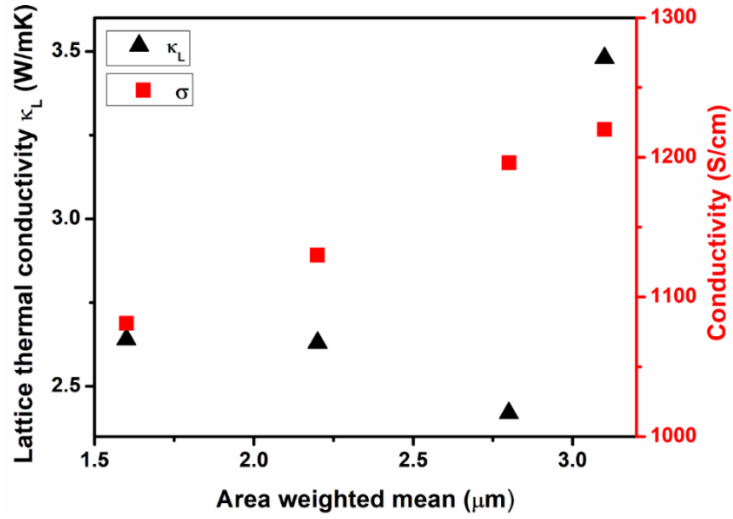

*Figure 10: Room temperature values of lattice thermal conductivity and electrical conductivity with area weighted mean values obtained from EBSD images*

*Table 1: The room temperature values of Lorenz Number (L), effective band mass (m\*), carrier concentration (n) and Hall mobility (μ) of $Hf_{0.3}Zr_{0.7}Co(Sn_{0.3}Sb_{0.7})_{1-x}Al_x$ with x=0%, 0.05%, 1%, 1.5% and 2.0%.*

| **Compositions** | **L** $(10^{-8}\ W\Omega/K^2)$ | **m\*** $(m_e)$ | **n** $(10^{21}\ cm^{-3})$ | **μ** $(cm^2V^{-1}s^{-1})$ |
|---|---|---|---|---|
| $Hf_{0.3}Zr_{0.7}CoSn_{0.3}Sb_{0.7}$ | 1.85 | 6.84 | 1.3 | 5.75 |
| $Hf_{0.3}Zr_{0.7}Co(Sn_{0.3}Sb_{0.7})_{0.995}Al_{0.005}$ | 1.84 | 8.59 | 1.37 | 5.45 |
| $Hf_{0.3}Zr_{0.7}Co(Sn_{0.3}Sb_{0.7})_{0.99}Al_{0.01}$ | 1.76 | 9.27 | 1.45 | 5.15 |
| $Hf_{0.3}Zr_{0.7}Co(Sn_{0.3}Sb_{0.7})_{0.985}Al_{0.015}$ | 1.76 | 9.45 | 1.52 | 4.64 |
| $Hf_{0.3}Zr_{0.7}Co(Sn_{0.3}Sb_{0.7})_{0.98}Al_{0.02}$ | 1.75 | 10.92 | 1.9 | 3.78 |

Room temperature Hall coefficients obtained for the samples listed in Table 1 were analyzed in order to complement first-principles calculations in elucidating the mechanism responsible for enhanced Seebeck coefficient. The room temperature values of carrier concentration and carrier mobility are summarized. Carrier mobility was calculated from room temperature resistivity measurements utilizing $\sigma=ne\mu$. To begin with, we analyze the experimental data in terms of a single parabolic band (SPB) model. The band structure within this model is governed by a single parameter, specifically the effective mass. This single parabolic band effective band mass (m\*) was obtained by solving the relevant equations (1), (2), and (3) for a parabolic band, with input from the room-temperature values of carrier concentration and Seebeck coefficient. A large effective band mass of ~10m\* was obtained. This is consistent with structure in the density of states near the band edge uncovered in first-principles calculations. The notable increase in effective band mass can reasonably account for the decrease in hole mobility and an increase in the Seebeck coefficient.

$$S = \pm \frac{k_b}{e} \left[ \frac{2F_1(\eta_F)}{F_0(\eta_F)} - \eta_F \right] \quad (1)$$

$$\text{where } F_n(\eta_F) = \frac{1}{\Gamma(n+1)} \int_0^\infty \frac{x^n}{1 + e^{(x-\eta_F)}} dx \quad (2)$$

$$n = \frac{4}{\pi} \left( \frac{2\pi m^* k_b T}{h^2} \right)^{\frac{3}{2}} F_{1/2}(\eta_F) \quad (3)$$

$\eta_F$ is the reduced Fermi energy and $F_n$ is the Fermi-Dirac integral.

However, it is typically the case that within a parabolic band model, while increased mass may be useful in obtaining an enhanced Seebeck coefficient, this comes at the price of decreased mobility and conductivity. It is noteworthy that the shape of the first principles density of states near the band edge does not closely follow the $E^{1/2}$ behavior of a parabolic band, where E is the distance between the Fermi level and the band edge. To proceed, we analyze the doping dependence of the Seebeck coefficient. Figure 11 shows a Pisarenko plot calculated using the inferred SPB effective mass for the measured undoped sample. This is using equation (4)[27].

$$S = \frac{8\pi^2 k_b^2 T}{3eh^2} m^* \left( \frac{\pi}{3n} \right)^{\frac{2}{3}} \quad (4)$$

where S is the Seebeck coefficient, $k_b$ is the Boltzmann constant, e is the electron charge, h is Planck's constant, T is temperature, *m\** the effective band mass, and n is the carrier concentration. As the carrier concentration increases, the Seebeck coefficient deviates significantly from the trend predicted by equation (4). This behavior implies that non-trivial electronic structure effects, in particular, a strong deviation from the parabolic band behavior upon which equation (4) is based. This is important because a very heavy effective mass m*~7 that would be consistent with the observed values of the Seebeck coefficient would not generally be consistent with the observed reasonable conductivity in these samples. This implies that the effective mass for the Seebeck coefficient is different from the transport effective mass for the conductivity. Thus, based on both transport measurements and first principles calculations, the electronic structure deviates from the parabolic band behavior, and this deviation enhances the Seebeck coefficient without the detrimental strong reduction in mobility that might otherwise be expected.

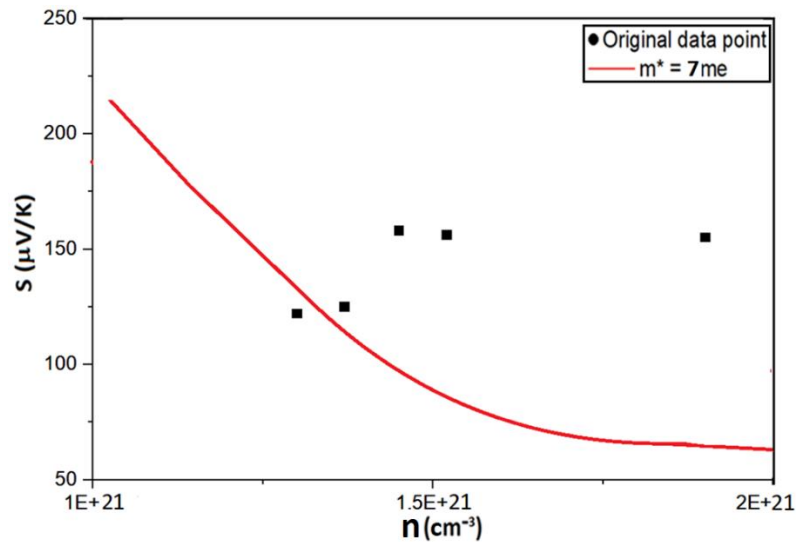

*Figure 11: Solid line is the Pisarenko plot showcasing the relationship between room temperature Seebeck coefficient, effective mass, and carrier concentration.*

## Conclusions

We have obtained a remarkable increase in ZT of a well-studied half-Heusler alloy from a previous maximum of 0.8 to 1.5 or higher. Surprisingly, a significant enhancement in thermoelectric properties can be achieved by applying as low as ~ 0.3 atomic percent aluminum (Al) doping. The results are attributed to the formation of resonant states near the valence band edge that modify the density of states near the Fermi energy, as confirmed from first principles calculations and complemented by electrical transport measurement. A moderate reduction in thermal conductivity is also observed, which is attributed to the unexpected formation of a highly heterogeneous microstructure that serves as an effective phonon scatter. Coupled with current high ZT n-type half-Heusler alloys in a similar material class, the present alloys are well suited for mid-to-high temperature thermal harvesting and power generation. Although the current approach originally focuses on enhancing the electrical components of materials, the unexpected heterogeneous microstructure due to minute doping nevertheless also leads to a moderate decrease in the lattice thermal conductivity. The success in achieving state-of-the-art thermoelectric properties through ordinary low-level doping in a well-studied alloy system is poised to spawn research activities on non-exotic thermoelectric materials.

## Conflict of Interest

The authors declare that they have no conflict of interest in the subject matter or materials discussed in this manuscript.

## Data availability statement

The raw/processed data required to reproduce these findings cannot be shared at this time as the data also forms part of an ongoing study.

# Conventional Half-Heusler Alloys Advance State-of-the-Art Thermoelectric Properties

Mousumi Mitra[1], Allen Benton[2], Md Sabbir Akhanda[3], Jie Qi[1], Mona Zebarjadi[3,4], David. J. Singh[5], Joseph Poon[1,4+]

[1]*Department of Physics, University of Virginia, Charlottesville, VA 22904*
[2]*Department of Physics & Astronomy, Clemson University, Clemson, SC 29631*
[3]*Department of Electrical Engineering, University of Virginia, Charlottesville, VA 22904*
[4]*Department of Material Science & Eng., University of Virginia, Charlottesville, VA 22904*
[5]*Department of Physics & Astronomy, University of Missouri, Columbia, MO 65211*

# Supplementary Materials

As described in the experimental section, two different sample series were synthesized, $(Hf_{0.3}Zr_{0.7})_{1-x}Al_xCoSn_{0.3}Sb_{0.7}$ and $Hf_{0.3}Zr_{0.7}Co_{1-x}Al_xSn_{0.3}Sb_{0.7}$. These compositions were doped with $Al_x$ (x = 0.01 and 0.015) on the Hf/Zr and Co site. As shown in Figure S1, Al saturates at a much lower doping level (x = 0.01) in these systems. The XRD patterns of these samples (Figure S1) show additional peaks (marked with star and triangle signs), which display distinct phase separation at the lower level of doping on those sites. According to the phase diagram of AlCo (bcc and B2 phases), Al and Co-rich impurity phases appear in these compositions, namely $Al_{16}Co_7$ and $Al_{1.4}Co_{0.6}$ as shown in Figure S1.

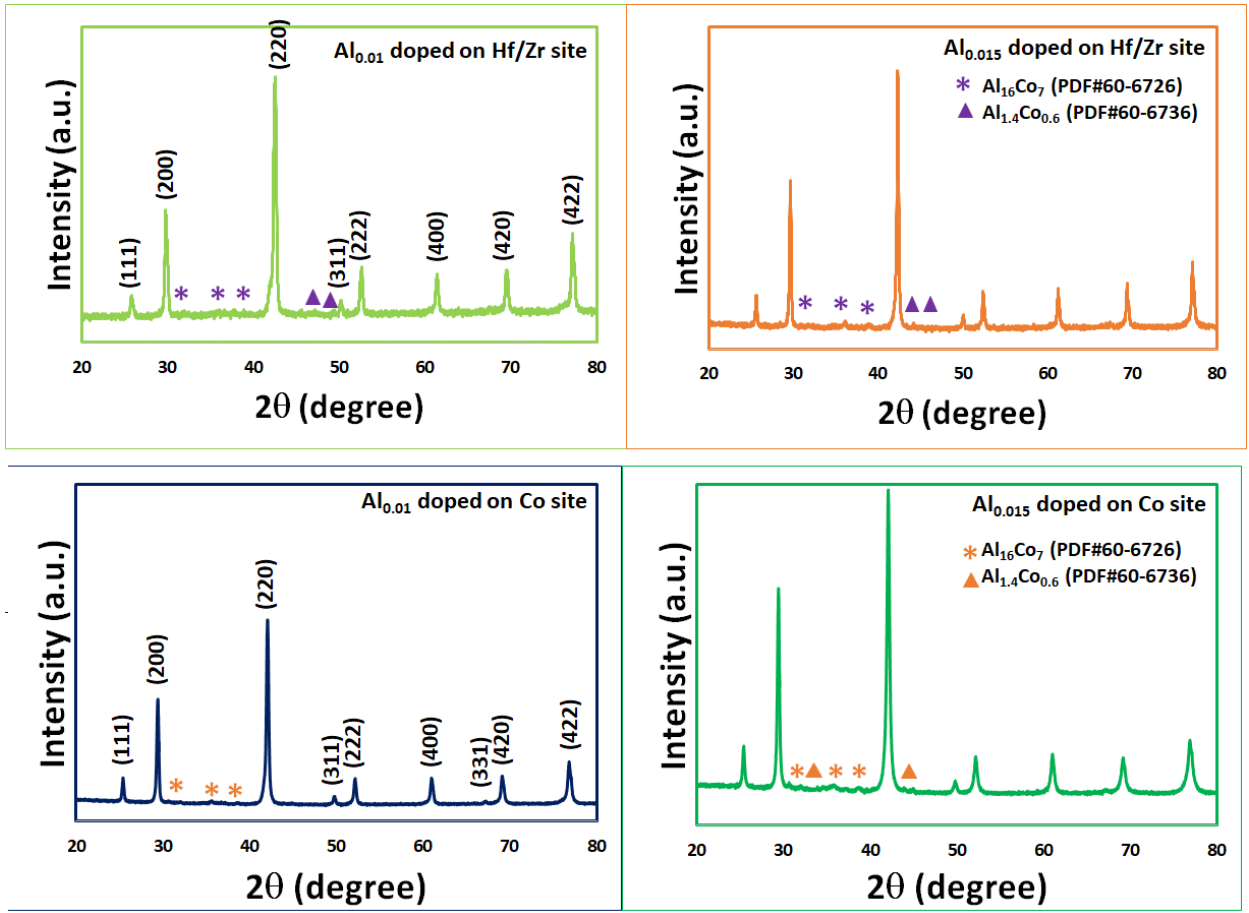

*Figure S1: Room temperature XRD spectra of $Hf_{0.3}Zr_{0.7}Co_{1-x}Al_xSn_{0.3}Sb_{0.7}$ and $(Hf_{0.3}Zr_{0.7})_{1-x}Al_xCo Sn_{0.3}Sb_{0.7}$. Peaks formed by secondary phases are marked with star and triangle signs.*

Figure S2 denotes the SEM and EDS line scan images of undoped ($Hf_{0.3}Zr_{0.7}CoSn_{0.3}Sb_{0.7}$), and $Al_{0.01}$ doped ($Hf_{0.3}Zr_{0.7}Co(Sn_{0.3}Sb_{0.7})_{0.99}Al_{0.01}$) samples. Figure S2c and S2d illustrate the compositional homogeneity in the undoped and $Al_{0.01}$ doped samples.

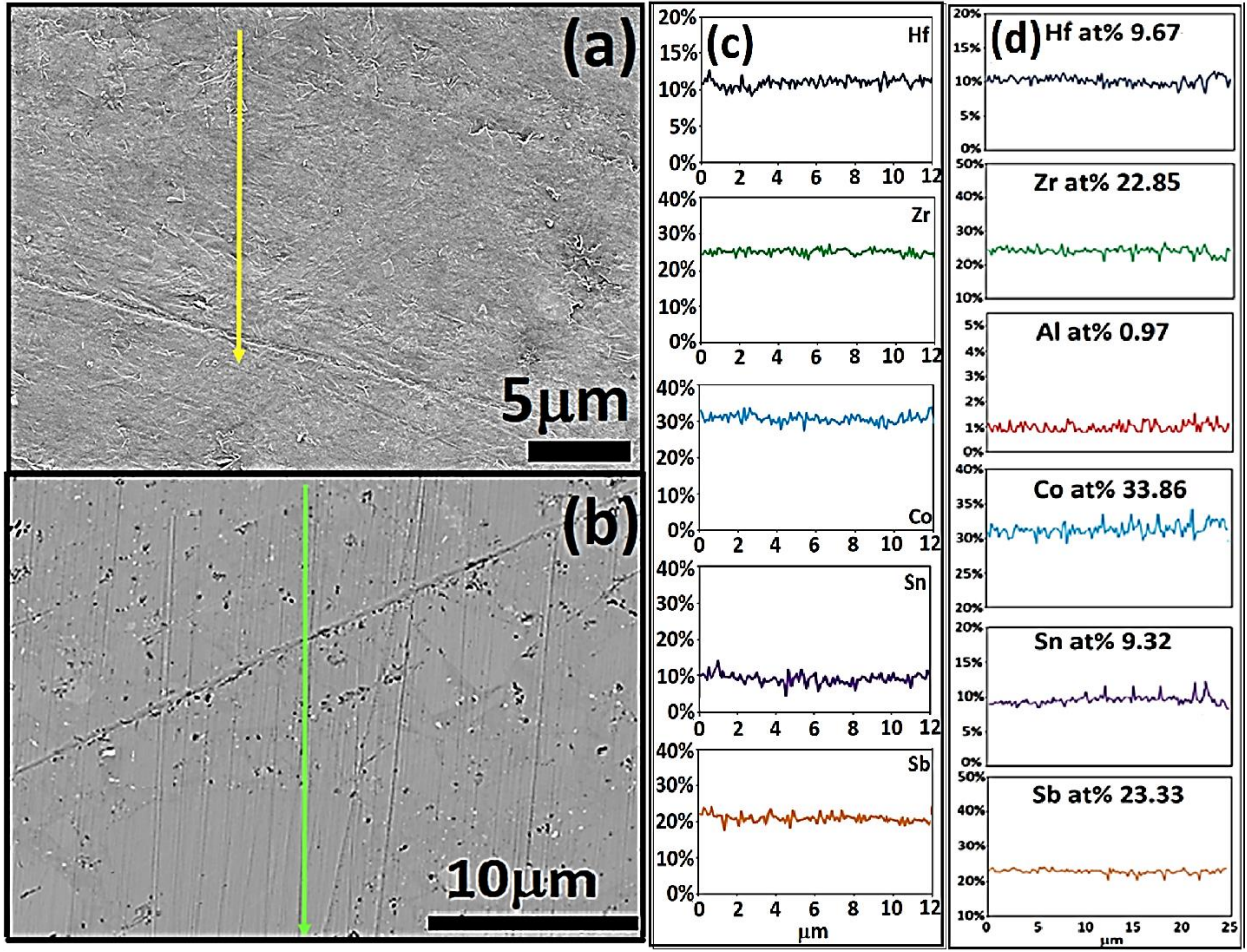

*Figure S2: SEM image of (a) $Hf_{0.3}Zr_{0.7}CoSn_{0.3}Sb_{0.7}$, and (b) $Hf_{0.3}Zr_{0.7}Co(Sn_{0.3}Sb_{0.7})_{0.99}Al_{0.01}$, EDS line scan of (c) $Hf_{0.3}Zr_{0.7}CoSn_{0.3}Sb_{0.7}$, and (d) $Hf_{0.3}Zr_{0.7}Co(Sn_{0.3}Sb_{0.7})_{0.99}Al_{0.01}$. Arrows indicate the location and direction of the abscissa in the corresponding EDS curves respectively.*

Figure S3a shows the SEM image of a typical region in the x = 0.02 sample. Figure S3b outlines the appearance of secondary phases in the Al0.02 doped sample. Figure S3c depicts the elemental mapping of the phase separated Al0.02 doped sample on the Sn/Sb site. According to EDS analysis (S3b), the phases were identified as half-Heusler $Hf_{0.3}Zr_{0.7}CoSn_{0.3}Sb_{0.7}$ (1), $Al_{16}Co_7$ (2), $Al_{1.4}Co_{0.6}$ (3), $Al_{13}Co_4$ (4), AlCo (5), Zr (6), Hf (7), Al (8) and Sn (9). The relative amount of all phases and their corresponding compositions are tabulated in supplementary Table S1.

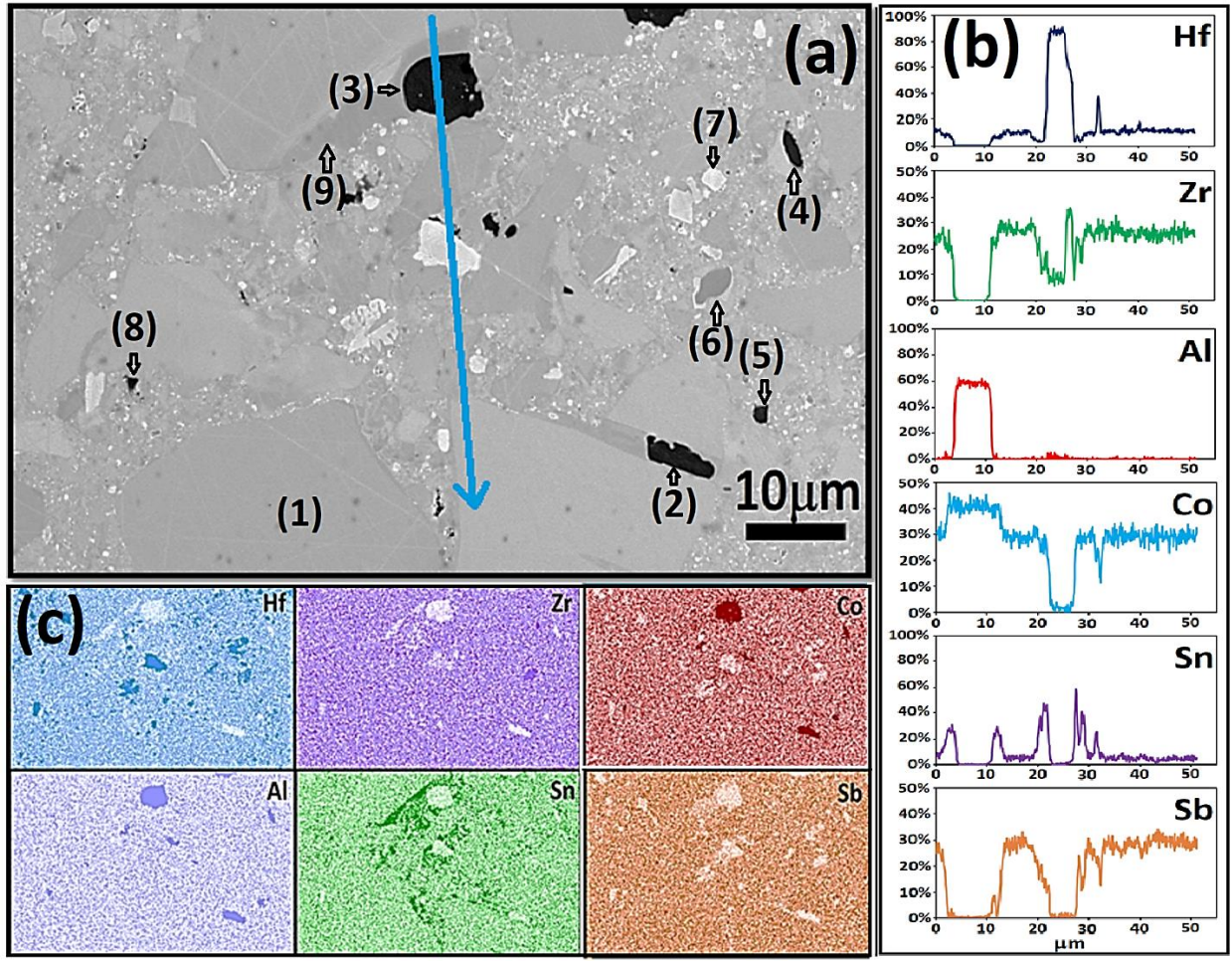

*Figure S3: (a) SEM image, (b) EDS line scan and (c) EDS elemental mapping of $Hf_{0.3}Zr_{0.7}Co(Sn_{0.3}Sb_{0.7})_{0.98}Al_{0.02}$ sample. Yellow arrow indicates the location and direction of the abscissa in the corresponding EDS curves (b). Legend of the phases for micrographs (a): half-Heusler $Hf_{0.3}Zr_{0.7}CoSn_{0.3}Sb_{0.7}$ (1), $Al_{16}Co_7$ (2), $Al_{1.4}Co_{0.6}$ (3), $Al_{13}Co_4$ (4), AlCo (5), Zr (6), Hf (7), Al (8) and Sn (9).*

*Supplementary Table 1: Different phases and relative quantity of all the phases of Al0.02 doped sample on the Sn/Sb site estimated from EDS curve:*

| Phases | Relative quantity, wt% |
|---|---|
| half-Heusler $Hf_{0.3}Zr_{0.7}CoSn_{0.3}Sb_{0.7}$ | 81 |
| $Al_{16}Co_7$ | 6 |
| $Al_{1.4}Co_{0.6}$ | 5 |
| $Al_{13}Co_4$ | 3 |
| AlCo | 2 |
| Zr | 0.3 |
| Hf | 2 |
| Al | 0.2 |
| Sn | 0.5 |

A closer look into the phase separation can be seen in the SEM images with the EDS line scan in figure S4. In S4A, the results for x = 0.01 doped on the Hf/Zr site indicate general inhomogeneity throughout. The small white grains appear to coincide with large peaks in Hf, and large dips for other elements indicate full separation of Hf from the lattice. Minimal variability is seen through other grains. Combining this with XRD data, we can conclude most grains are some forms of half-Heusler structure with various degrees of separation between elemental compositions. Similarly, Figure S4B, highlights compositions of the x = 0.01 on the Co site sample. The small dark grain in the image corresponds to the large peak in Al and Co, representing some form of $Al_xCo_y$ forming. We can again combine XRD results and EDS to conclude the main phase is some form of half-Heusler structure; however, the Al content is more concentrated in the darker grains in this case. We find the Al content is less homogeneous when doped to the Hf/Zr and Co sites versus the Sn/Sb site. Large amounts of Al are found forming alternate phases outside the lattice, while we found a much higher level of homogeneity in the sample doped on the Sn/Sb site.

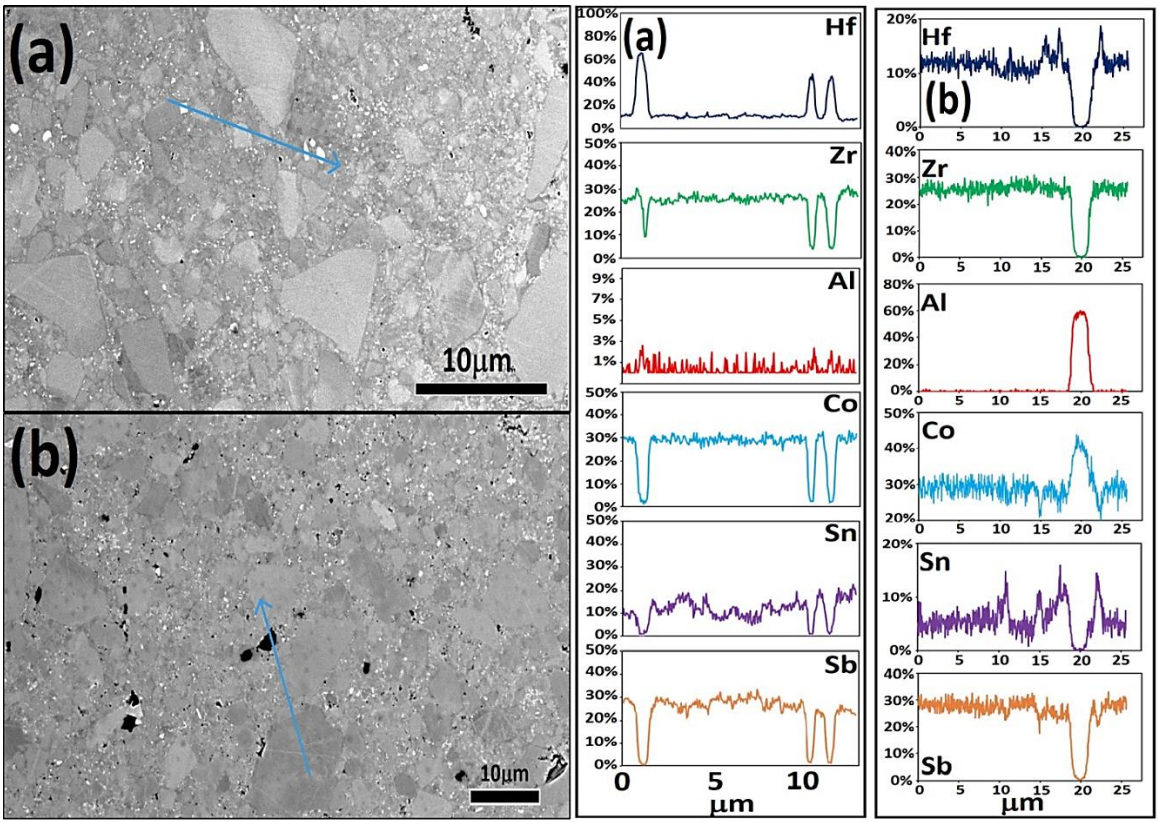

*Figure S4: SEM and EDS line scan of (a) $Hf_{0.3}Zr_{0.7}Co_{0.09}Al_{0.01}Sn_{0.3}Sb_{0.7}$ and (b) $(Hf_{0.3}Zr_{0.7})_{0.09}Al_{0.01}CoSn_{0.3}Sb_{0.7}$. Blue arrows indicate the location and direction of the abscissa in the corresponding EDS curves.*

Figure S5 shows the resistivity (S5A), Seebeck coefficient (S5B), and power factor (S5C) plots for Al0.01 and Al0.015 doped on Hf/Zr and Co site samples. It has been clear from the plots that due to the emergence of the second phase for the Al0.015 doped on Hf/Zr and Co site samples, the power factor values are diminished compared to Al0.01 doped samples for the whole temperature range from 300-1100 K.

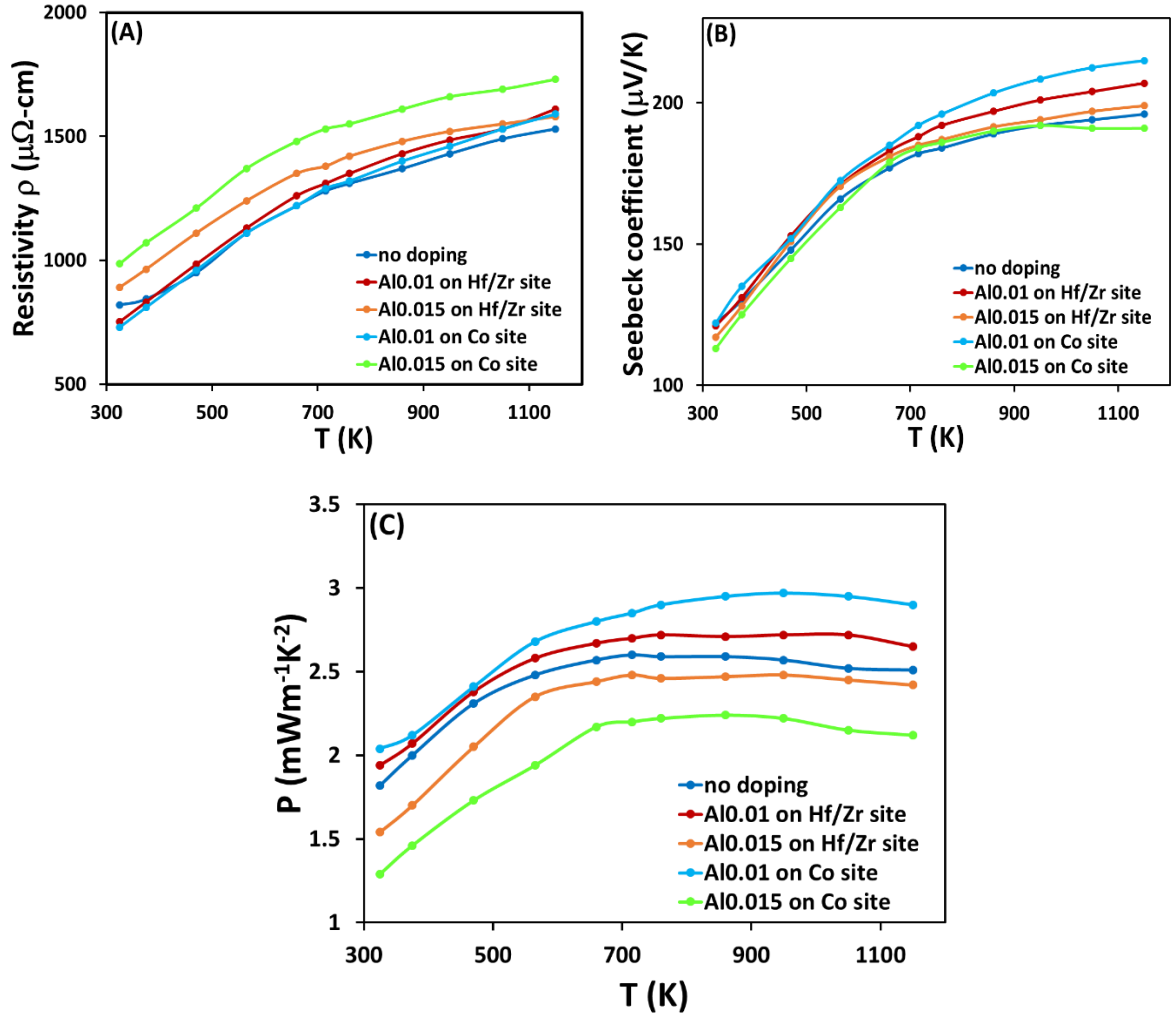

*Figure S5: (A) Electrical resistivity, (B) Seebeck coefficient and (C) Power factor curves for $Hf_{0.3}Zr_{0.7}Co\ Sn_{0.3}Sb_{0.7}$ (dark blue), $(Hf_{0.3}Zr_{0.7})_{0.085}Al_{0.015}CoSn_{0.3}Sb_{0.7}$ (orange), $Hf_{0.3}Zr_{0.7}Co_{0.085}Al_{0.015}Sn_{0.3}Sb_{0.7}$ (green), $(Hf_{0.3}Zr_{0.7})_{0.09}Al_{0.01}CoSn_{0.3}Sb_{0.7}$ (red), and $Hf_{0.3}Zr_{0.7}Co_{0.09}Al_{0.01}Sn_{0.3}Sb_{0.7}$ (light blue).*